\documentstyle[preprint,aps]{revtex}

\newcommand{\la}{\langle}
\newcommand{\ra}{\rangle}
\newcommand{\beq}{\begin{eqnarray}}
\newcommand{\eeq}{\end{eqnarray}}

\newcommand{\cl}{\centerline}

\tighten

\begin{document}

\draft

\title{  Light Vector Mesons at Finite Baryon Density
\footnote{Talk presented at RHIC Summer Study '96, Brookhaven
 Theory Workshop on Relativistic Heavy Ions, (BNL, July 8-19, 1996).}}

\author{ T.\ Hatsuda}

\address{Institute of Physics, University of Tsukuba,
 Tsukuba, Ibaraki 305, Japan}

\maketitle

\begin{abstract}
We summarize the current theoretical and experimental status
of the spectral changes of  vector mesons ($\rho$, $\omega$, $\phi$)
 at finite baryon density.
  Various approaches including QCD sum rules, effective 
 theory of hadrons and bag models  show
 decreasing of the vector meson masses in nuclear matter.
  Possibility to detect  
 the mass shift through lepton pairs
  in $\gamma-A$, $p-A$ and $A-A$ reactions are also discussed.

\end{abstract}

\setcounter{footnote}{0}
\renewcommand{\thefootnote}{\alph{footnote}}

\section{Introduction}

 At high temperature ($T$) and density ($\rho$),
  hadronic matter is expected to undergo  a phase  
  transition to the
  quark-gluon plasma.
 The order parameter characterizing the transition 
 is the chiral quark condensate $\langle \bar{q}q \rangle$, the 
 absolute value of which decreases as ($T$,$\rho$) increases.
  Numerical simulations of quantum chromodynamics (QCD)
 on the lattice 
  are actively pursued to determine the 
  precise nature of the transition at finite $T$ \cite{ukawa} 
 and various model calculations  have been done
 to look for the observable signature of the phase transition.
 
In this talk, I  will concentrate on one of the 
 interesting critical phenomena  associated with the 
 QCD phase transition, namely the spectral change of hadrons,
 in particular the mass shift of  light vector-mesons ($\rho$, $\omega$
 and $\phi$) in nuclear matter at zero $T$.
 The vector mesons are unique in the sense that
  they decay into lepton
 pairs ($e^+ e^-$ and $\mu^+ \mu^-$) which can be detected experimentally
 without much disturbance by  complicated hadronic interactions.
 
 In section 2, I will review the current knowledge of the 
 quark condensate in medium. In section 3, various approaches to
 calculate the vector meson masses in nuclear matter are summarized.
 Experimental possibilities to detect the spectral change 
 are discussed in section 4.  Concluding remarks are given in section 5.

\section{Quark condensates in nuclear matter}

  The medium modification of the quark condensate
 has been calculated since then 
 by various methods (lattice QCD, chiral perturbation
 theory, Nambu-Jona-Lasinio model etc). See an overview \cite{review}.
 By these studies, it turned out that
  there is one noticeable difference between the
 behavior of $\langle \bar{q} q \rangle$ at finite $T$ (with $\rho =0$)
  and that at finite
 $\rho$ (with $T=0$):  In the former case, the significant change of
 the condensate can be seen only near the critical point
 $T\sim T_c$ \cite{GL}.  On the other hand,  in the
 latter case, 
 $O(30 \%)$ change of $\langle \bar{q} q \rangle$
 could be seen even in normal nuclear-matter density. This 
 observation is based on the following formula in the
 fermi-gas approximation (independent particle approximation)\cite{DL}
\begin{eqnarray}
\label{condrho}
{\langle \bar{u}u \rangle \over \langle \bar{u}u \rangle_{0}}
  \simeq  1- {4\Sigma_{\pi N}\over f_{\pi}^2m_{\pi}^2}
 \int^{p_{_F}} {d^3p \over (2\pi)^3} {M_N \over E(p)} \ \ . 
\end{eqnarray}
Here $m_N (m_{\pi})$ is the nucleon (pion) mass, $f_{\pi}$ is the pion decay 
 constant,  $\Sigma_{\pi N} = (45 \pm 10) $MeV is the $\pi N$ sigma term,
 and $E(p) \equiv \sqrt{p^2 + M_N^2}$.  $\langle \cdot \rangle $
 and $\langle \cdot \rangle_{0}$ denote the expectation value
 in nuclear matter and that in the vacuum respectively.
 The integration for the nucleon momentum $p$ should be taken
  from 0 to the
 fermi momentum  $p_{_F}$. 
 At normal nuclear matter density ($\rho=\rho_0=0.17/{\rm fm}^3$), 
 the above formula gives (34$\pm$8)\% reduction of the
 chiral condensate from the vacuum value. 
 In Fig.1, 
${\langle \bar{u}u \rangle / \langle \bar{u}u \rangle_{0}}$ as
 well as the strangeness condensate 
$ {\langle \bar{s}s \rangle / \langle \bar{s}s \rangle_{0}}$
  are shown in the linear density approximation 
 \cite{QM91}, where the uncertainty of $\Sigma_{\pi N}$
  is considered. 
 Estimates taking into account the fermi motion and 
 the nuclear correlatons show that these
 corrections at $\rho = \rho_0$ 
 are less than the above uncertainty  \cite{Ko94}.
 
 Unfortunately, the condensate itself is not a direct observable
 and one has to look for physical quantities which are measurable and
 simultaneously  
 sensitive to the change of the condensate. 
 The masses of light vector-mesons are the leading candidates
 of such quantities.

\section{Vector mesons in nuclear matter}

Let's   consider $\rho$, $\omega$ and $\phi$  mesons propagating inside
 the nuclear matter.
 Adopting the same fermi-gas approximation with (\ref{condrho}) and taking
 the vector meson at rest (${\bf q}=0$),
 one can generally write the mass-squared shift  as
\begin{eqnarray} 
\label{massshiftf}
\delta m^2_{_V} \equiv m^{*2}_{_V} - m^2_{_V} = 4 \int^{p_F}
  {d^3 p \over (2 \pi)^3  }  {M_{_N} \over E(p)}  f_{VN}({\bf p}), 
\end{eqnarray}
where $f_{VN}({\bf p})$ denotes the vector-meson (V) -- nucleon (N)
 forward scattering amplitude 
 in the relativistic normalization, and $m_V^* (m_V)$ denotes the 
 vector meson mass in nuclear matter (vacuum).
 Here, we took spin-isospin average for the nucleon states
 in $f_{VN}$.
 If one can calculate $f_{VN}({\bf p})$ reasonably well in the range
 $0 < p < p_{_F}=270$ MeV
 (or $1709\ {\rm MeV} < \sqrt{s} < 1726\ {\rm MeV} $ in terms of the
 $V-N$ invariant mass), one can predict the mass shift.
 Unfortunately, this is a difficult task:
  $f_{VN}({\bf p})$ is not a  constant  
  in the above range since there are at least two 
 s-channel resonances $N(1710), N(1720)$ in the above interval
 and two nearby resonances $N(1700)$ and $\Delta(1700)$.
  They all couple to the $\rho-N$ system \cite{PDG}
 and give  variation
 of $f_{VN}({\bf p})$ as a function of $p$ in principle.
 From this reason,  one should develop
   other methods to estimate
 $\delta m_V^2$ without
 refering to the detailed form of $f_{VN}({\bf p})$.
 We will briefly review two
of such approaches
  in the following subsections, namely the QCD sum rules
 and  effective theories of hadron.

\subsection{QCD sum rules}

 The  QCD sum rules (QSR) for vector mesons in nuclear matter
 were first developed by Hatsuda and Lee \cite{HL}.
 In their approach,  one
 starts with  the retarded current correlation function in 
nuclear matter,
\begin{eqnarray}
\label{correlator}
\Pi_{\mu \nu} (\omega , {\bf q})
=i \int d^4x e^{iqx}  \langle {\rm R} J_{\mu}(x) J_{\nu}(0) \rangle \ \ ,
\end{eqnarray}
where  $q^\mu \equiv (\omega , {\bf q})$ and 
 ${\rm R} J_{\mu}(x) J_{\nu}(0) \equiv 
  \theta(x^0) [J_{\mu}(x), J_{\nu}(0)]$ with the source
currents $J_\mu$ defined as $J_\mu^{\rho,\omega} 
={1\over2}(\bar{u}\gamma_\mu u \mp \bar{d}\gamma_\mu d)$ 
($- (+)$ is for the $\rho^0 (\omega)$-meson) and
 $J_{\mu}^{\phi} = \ \bar{s} \gamma_{\mu} s$.
  Although there are two independent
invariants in medium (transverse and longitudinal polarization),
 they coincide  in the limit  ${\bf q} \rightarrow 0$
 and reduce to
 $\Pi_{\mu \mu}/(-3\omega^2) \equiv \Pi$.
 $\Pi $ satisfies the following dispersion
relation,
\begin{eqnarray}
\label{dispersion2}
{\rm Re} \Pi (\omega^2) =
 {1 \over \pi} {\rm P} \int_0^{\infty} du^2
{ {\rm Im} \Pi(u) \over u^2-\omega^2} + ({\rm subtraction}). 
\end{eqnarray}
In QSR, the spectral density ${\rm Im} \Pi$
 is  modeled with several 
phenomenological parameters, while  ${\rm Re} \Pi$
  is calculated using the 
operator product expansion (OPE).
 The phenomenological parameters are then  extracted  
by matching the left and right hand side of (\ref{dispersion2})
 in the asymptotic region $\omega^2 \rightarrow - \infty $.
 The density dependence in the OPE side is solely determined by the
 density dependent condensates which are evaluated from
 low energy theorems or from the parton distribution of the 
  nucleon \cite{HL}.

 In the medium,  we have three kinds of structure in the spectral density:
 the resonance poles,
  the continuum and the Landau damping contribution.
 For ${\bf q} \rightarrow 0$, the last contribution 
 is calculable {\em exactly} 
 and behaves like  a pole at $\omega^2=0$ \cite{HL,BS}.
  In total,  the hadronic spectral function looks as
\begin{eqnarray}
\label{phen}
 8 \pi {\rm Im} \Pi(u > 0^-)  =  
 \delta(u^2) \rho_{sc}+F^* \delta(u^2-m_V^{2*})
+(1+\frac{\alpha_s}{\pi})\theta(u^2-S_0^*) 
  \equiv  \rho_{had.}(u^2) ,
\end{eqnarray}
with $\rho_{sc} = 2 \pi^2 \rho /\sqrt{p_F^2 + M_N^2} \simeq
 2 \pi^2 \rho /M_N$.  $m_V^*$, $F^*$ and $S_0^*$ 
 are the three phenomenological
 parameters in nuclear matter to be determined by the sum rules.

Matching the OPE side 
 and the phenomenological side via the dispersion relation 
in the asymptotic region $\omega^2 \rightarrow - \infty$,
 we can 
  relate the resonance  parameters to the density dependent condensates.  
 There are two major procedures for this matching, namely 
 the Borel sum rules (BSR) \cite{SVZ} and the
  finite energy sum rules (FESR) \cite{PIV},
 which can be summarized as
\begin{eqnarray}
\label{sumrules}
\int_0^{\infty}& ds\ W(s)& \ [\rho_{had.}(s) - \rho_{_{OPE}}(s) ]  =0 ,\\ 
  & W(s)  =&  s^n \ \theta(S_0 -s)  \ \ \  ({\rm  FESR}),
   \ \ \ \ \ \ \ e^{-s/M^2}  \ \ \  ({\rm \ BSR}).
\end{eqnarray}
Here the spectral function $\rho_{had.}(s)$ stands for eq.(\ref{phen}).
 $\rho_{_{OPE}}(s)$ is a hypothetical imaginary part of
 $\Pi$ obtained from OPE.

 To make  quantitative analyses of spectral 
 parameters, the stability analysis
 based on the  Borel transform is more suitable than FESR.
 Since the Borel mass $M$ is a fictitious parameter introduced 
 in the sum rule, the physical quantities should be insensitive to
 the change of $M$ within a Borel interval $M_{\min} < M < M_{\max}$;
 namely the principle of minimum sensitivity (PMS) is used.
 One can accomplish this insensitivity by choosing
 $S_0^*$ suitably at given density.
 In Fig. 2, the  Borel curves for the $\rho (\omega)$ meson
 for three different values of baryon density are shown with $S_0^*$
 chosen to make the Borel curve as flat as possible in the interval  
 $ 0.41 {\rm GeV}^2 < M^2 < 1.30 {\rm GeV}^2$. The upper (lower)
 bound of the Borel interval is determined so that the power (continuum)
 correction after the Borel tranform does not exceed 30 $\%$ of 
 the lowest order term in OPE.

  By making a linear fit of the result,  one obtains \cite{HL}
\begin{eqnarray}
\label{mass-shift}
{m_{\rho,\omega}^* \over m_{\rho,\omega}} & = &  1- (0.16 \pm 0.06) 
{\rho \over \rho_0}, \\ 
\label{threshold-shift}
\sqrt{{S_0^* \over S_0}} & = & 1- (0.15 \pm 0.05)
 {\rho \over \rho_0}, \\ 
\label{F-shift}
{F^* \over F} &  = & 1- (0.24 \pm 0.07) {\rho \over \rho_0},  
\end{eqnarray}
and 
\begin{eqnarray}
\label{mass-shift2}
{m_{\phi}^* \over m_{\phi}}  =  1- (0.15 \pm 0.05)\  y\  
{\rho \over \rho_0},
\end{eqnarray}
where $y$ is the OZI breaking parameter in QCD defined
 as $y=2\la \bar{s}s \ra_N/ \la \bar{u}u + \bar{d} d \ra_N$
 with $\la \cdot \ra_N$ being the nucleon matrix element.
 $y$ takes the value
 $0.1 - 0.2$ \cite{HL}.  
  The decrease in eqs. (\ref{mass-shift},\ref{mass-shift2}) 
 is dictated by the density dependent
  condensates $\langle \bar{q} q \rangle$,  $\langle (\bar{q} q)^2 \rangle$
 and  $\langle \bar{q} \gamma_{\mu} D_{\nu} q \rangle$.
 The errors in the above formulas are originating from the uncertainties
 of the density dependence of the these condensates.
 The contribution of the
  quark-gluon mixed operator with twist 4, \cite{HL} which may
   possibly weaken the mass shift, is neglected in the above.
   Shown in Fig.3 is the mass shift given in 
 eqs. (\ref{mass-shift},\ref{mass-shift2}) with possible 
 theoretical uncertainties.

 Some sophistications of the QSR analyses by Hatsuda and Lee
 have been done later by several authors.

\noindent
(i) Asakawa and Ko have introduced  a more realistic
 spectral function  than (\ref{phen})  by taking into account
 the width of the $\rho$-meson and the effect of the collisional
 broadening due to the   $\pi - N - \Delta -\rho$
 dynamics \cite{ASKO}.  By doing the similar QSR analysis as above,
 they found that the negative mass shift 
  persists even in this realistic case.  The width of the
 rho meson in their calculation  decreases as density increases, which 
implies that the phase space suppression from the $\rho \rightarrow 2 \pi$
 process overcomes the collisional broadening at finite density.
 Further examination of this interplay between the mass shift and
 the collisional broadening is important in relation to the future
 experiments. Also, finite temperature generalization of the Asakawa-Ko's
 calculation should be  done.

\noindent
(ii) Monte Calro based error analysis was applied to the 
 Borel sum rule by Jin and Leinweber \cite{JL} instead of the
 Borel stability or PMS analysis employed in \cite{HL}.
 They found 
 $ m_{\rho,\omega}^*/ m_{\rho,\omega} =  1- (0.22 \pm 0.08)  
(\rho /\rho_0)$ and 
 $m_{\phi}^* /m_{\phi} =  1- (0.01 \pm 0.01) 
(\rho / \rho_0)$, which are consistent with
 eqs. (\ref{mass-shift},\ref{mass-shift2}) within 
 the error bars.

\noindent
(iii) Koike analysed an {\em effective} scattering amplitude $\bar{f}_{VN}$
 defined as   $\delta m^2_{_V} \equiv {\bar f}_{VN} \cdot \rho $ 
 using the QSR in the vacuum \cite{Koike}.  Although his original calculation
 predicting ${\bar f}_{VN} > 0$ is in error as was pointed out in
 ref.\cite{HL,JL}, revised calculation gives a consistent result with
 eqs. (\ref{mass-shift},\ref{mass-shift2}) within the error bars \cite{Koike2}.
 Note here that ${\bar f}_{VN}$ does not have 
  direct relation to the scattering length at zero
 momentum $f_{VN}(0)$.

\subsection{Effective theories}

 There have been  many attempts so far to calculate the 
 spectral change of the vector mesons using effective theories of QCD.
 The first attempt by Chin \cite{chin}
 using the quantum hadrodynamics (QHD) 
 shows increasing $\omega$-meson mass
 in medium due to a process  analogous to the Compton scattering;
\begin{eqnarray}
 \omega + N 
  \rightarrow \omega + N.
\end{eqnarray}
  For the $\rho$-meson, similar but more sophisticated calculations
 taking into account  $\Delta$-resonance and in-medium pion
 show a slight increase of the $\rho$-meson mass \cite{herman}.
  In these calculations, only the 
  polarization of the Fermi sea (the particle-hole excitations)
 was considered. Also their predictions are
 different from the general assertion by Brown and Rho claiming
 that all the hadron masses except for pion should decrease \cite{BR}.

  On the other hand, Saito, Maruyama and Soutome, and 
  Kurasawa and Suzuki \cite{KS}
   have realized that 
  the mass of the $\omega$-meson is affected substantially
 by the vacuum polarization of the nucleon in medium 
\begin{eqnarray}
 \omega \rightarrow N^* \bar{N}^* \rightarrow \omega,
\end{eqnarray}
 where  $N^*$ is the nucleon
  in nuclear matter which has smaller effective mass than that in the vacuum.
  They show  that the vacuum polarization  dominates over the 
 Fermi-sea polarization in QHD  and leads decreasing
 vector meson mass. This conclusion was later confirmed by several
 groups \cite{PW} and was generalized for the $\rho$ and $\phi$ mesons
  \cite{HS}.
  Jaminon and Ripka has also reached a similar conclusion
  by using a model of vector mesons coupled to  constituent quarks
  \cite{JR}.
 
  Saito and Thomas have examined a rather different 
 but comprehensive model (bag model combined with QHD)
 and 
 found  decreasing vector-meson masses \cite{ST};
 $ m_{\rho,\omega}^* / m_{\rho,\omega}  \sim   1- 0.09  
(\rho / \rho_0)$.
  The spectral shift 
 of the quarks inside the bag induced by the existence of 
 nuclear medium plays a key role in this approach. 

   Basic idea common in the approaches predicting the
 decreasing mass may be summarized as follows.
 In nuclear matter, scalar ($\sigma$) and vector ($\omega$)
 mean-fields are induced by the  nucleon sources.
  These mean-fields give back-reactions
 to the nucleon propagation in nuclear matter and modify
  its self-energy. This is an origin of the effective nucleon mass
 $M_N^* < M_N$ in  the relativistic
 models for nuclear matter.
  The same mean-fields should also affect the propagation of 
  vector mesons in nuclear medium.
 In QSR, the quark condensates act on the quark propagator
 as density dependent mean-fields.
  In QHD, the coupling of the mean-field with the vector
 mesons are taken into account through the short distant
  nucleon
  loop with the effective mass $M_N^*$.
  In the bag-model, the mean fields outside the bag
 acts on quarks confined in the 
 bag and change their energy spectrum. 
 
 Let us show here that one can understand the 
 negative mass shift of the vector mesons in a simple
 and intuitive  way in the context of QHD.
 More quantitative discussion will be given in the later section.
  After renormalizing infinities in the vacuum loop,
  the density-dependent part of the
 Dirac-sea polarization
 to the vector-meson propagator is approximately  written as 
\begin{eqnarray}
\label{Zfactor}
D(q) \simeq {1 \over Z^{-1} q^2 - m_V^2} = {Z \over q^2 - Z m_V^2} ,
\end{eqnarray}
where $Z$ being the finite wave-function renormalization constant
 in medium.  The pole position is thus obtained as $m_V^* = \sqrt{Z} m_V$.
 Because of the current conservation, only the
 wave function part of the propagator is modified in medium.
 Since the effective mass of the nucleon decreases in medium
 ($M_N^*/M_N < 1$), physical 
 vector mesons have more probability to be in
 virtual baryon$-$ anti-baryon
  pairs compared to that in the vacuum.  This means
 $Z < 1$, which leads to $m_V^* /m_V \equiv Z  < 1$ \cite{PW,HS}.
\begin{eqnarray}
M_N^*/M_N < 1 \ \ \  \rightarrow \ \ \  Z < 1 \ \ \  \rightarrow \ \ \ 
  m_V^*/m_V \equiv Z   < 1 \ \ \ . 
\end{eqnarray}

\section{Experiments}

How one can detect the spectral change of vector mesons  in experiments?
 One of the promising ideas 
 is to use heavy nuclei and produce vector mesons in 
  $\gamma -A$ or $p-A$ reactions.
 Suppose one could create a vector meson at the center
 of a heavy nucleous. (It does not matter whether it is created  at the
 nuclear surface or at the center as far as the produced
 vector mesons run through the nucleous before the hadronic decay).
It is easy to see that the number of lepton pairs decaying inside the
 nucleous $N_{in}(l^+l^-)$ and that outside the nucleous
 $N_{out}(l^+l^-)$ are related as
\begin{eqnarray}
\label{NN}
{N_{in}(l^+l^-) \over N_{out}(l^+l^-)}  \sim  
{ 1 - e^{- \Gamma_{tot} R} \over e^{- \Gamma_{tot}R} } \ \ ,
\end{eqnarray}
where $\Gamma_{tot}$ denotes the total width of vector mesons
 ((1.3fm)$^{-1}$, (23fm)$^{-1}$ and (45fm)$^{-1}$ for $\rho$, $\omega$ and
 $\phi$, respectively) and $R$ being the nuclear radius.
 Eq.(\ref{NN}) shows that even the $\phi$ meson has considerable
 fraction of $N_{in}/N_{out}$ if the target nucleous is big enough.

 There exist already some proposals to look for the mass shift of 
 vector mesons in nuclear medium  \cite{SE}.
 One is by Shimizu et al. who  propose an experiment
 to  create $\rho$ and $\omega$
 in heavy nuclei using coherent photon - nucleus reaction and
  subsequently  detect the lepton pairs from $\rho$ and $\omega$.
 Enyo et al. propose to create $\phi$ meson in heavy nuclei
 using the proton-nucleus reaction and 
 to measure kaon pairs as well as the lepton pairs.
 By doing this, one can study not only the mass shift 
but also the change of the leptonic vs hadronic branching ratio
$r = \Gamma(\phi \rightarrow e^+ e^- ) /\Gamma (\phi \rightarrow K^+ K^- )$.
 Since $m_{\phi}$ is very close to $2m_{K}$ in the vacuum,
 any modification of the $\phi$-mass or the $K$-mass changes
 the ratio $r$ substantially as a function of mass number of 
 the target nucleous.
  Similar kinds of experiments are also planned
 at GSI.

 There are also on-going  heavy ion experiments at SPS (CERN) and AGS (BNL)
 where  high density matter is likely to be formed.
 In particular, CERES/NA45 and HELIOS-3 at CERN
   reported enhancement of the
  lepton pairs below the $\rho$ resonance \cite{ceres,helios},
  which may not be explained
 by the conventional sources of lepton pairs.
 If this phenomena is real, low mass enhancement  of the lepton pair
  spectrum
 expected by the mass shift of the vector mesons could be a
  possible explanation \cite{ceres2}.
  In nuclear collisions at  higher energies (RHIC and LHC),
  hot hadronic matter or possibly the quark-gluon-plasma 
 with low baryon density are expected to be formed.  In such cases,
 double $\phi$-peak structure proposed by Asakawa and Ko \cite{theory}
 as well as  the spectral change of $\rho$, $\omega$ and scalar
 mesons \cite{HKL}
will be a distinct signal of the chiral restoration in QCD.

\section{Concluding remarks}

The spectral change of the elementary excitations in medium
 is an exciting new possibility in QCD.  
 By studying such phenomenon, one can learn the structure
 of the hadrons and the QCD ground state at finite $(T, \rho$)
 simultaneously.   Theoretical approaches such as the QCD sum rules
 and the hadronic effective theories 
 predict that the  light vector mesons ($\rho$, $\omega$ and $\phi$) 
 are sensitive to the partial restoration of chiral symmetry
 in hot/dense medium.  These mesons are good 
 probes experimentally, since they decay into lepton pairs which penetrate
 the hadronic medium without loosing much information.
  Thus, the lepton pair spectroscopy in QCD will tell us a 
 lot about the detailed structure of the hot/dense matter, which
 is quite similar to the soft-mode spectroscopy
 by the photon and neutron scattering experiments in solid state physics.

\vspace{1cm}

\cl{{\bf Acknowledgements}}

\vspace{0.4cm}

This work was supported by the Grants-in-Aids of the Japanese Ministry
 of Education, Science and Culture (No. 06102004). 

\vspace{2cm}

\centerline{{\bf Figure Captions}}
\begin{description} 

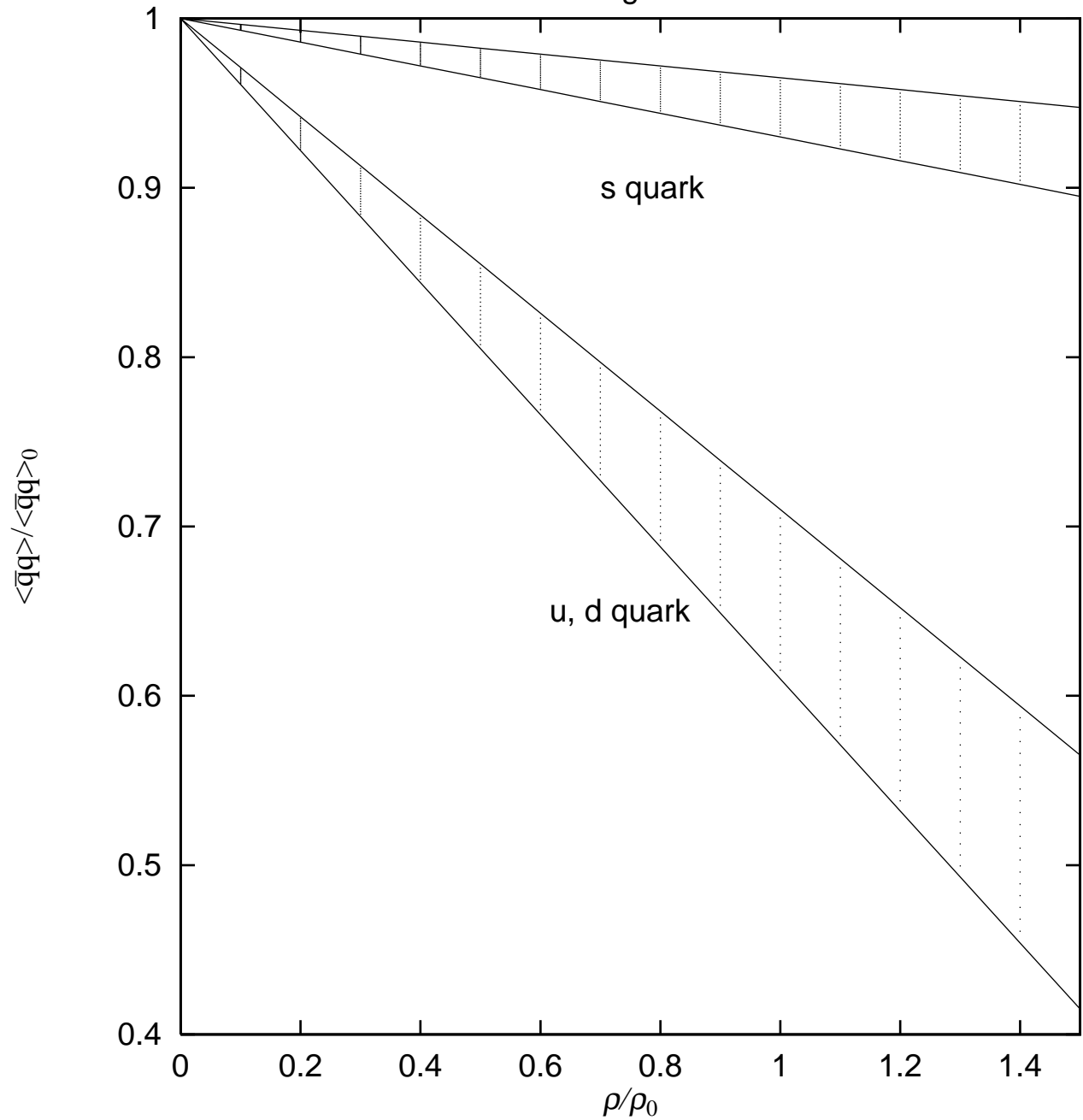
\item [Fig.1]
The light quark condensates in N=Z nuclear matter in the
 linear density approximation.
  Theoretical uncertainty of the $\pi N$ sigma term is
  taken into account. We take $y = 0.12$  for 
  the OZI breaking parameter, where 
 $y \equiv 2\la \bar{s}s \ra_N/ \la \bar{u}u + \bar{d} d \ra_N$
 with $\la \cdot \ra_N$ being the nucleon matrix element.

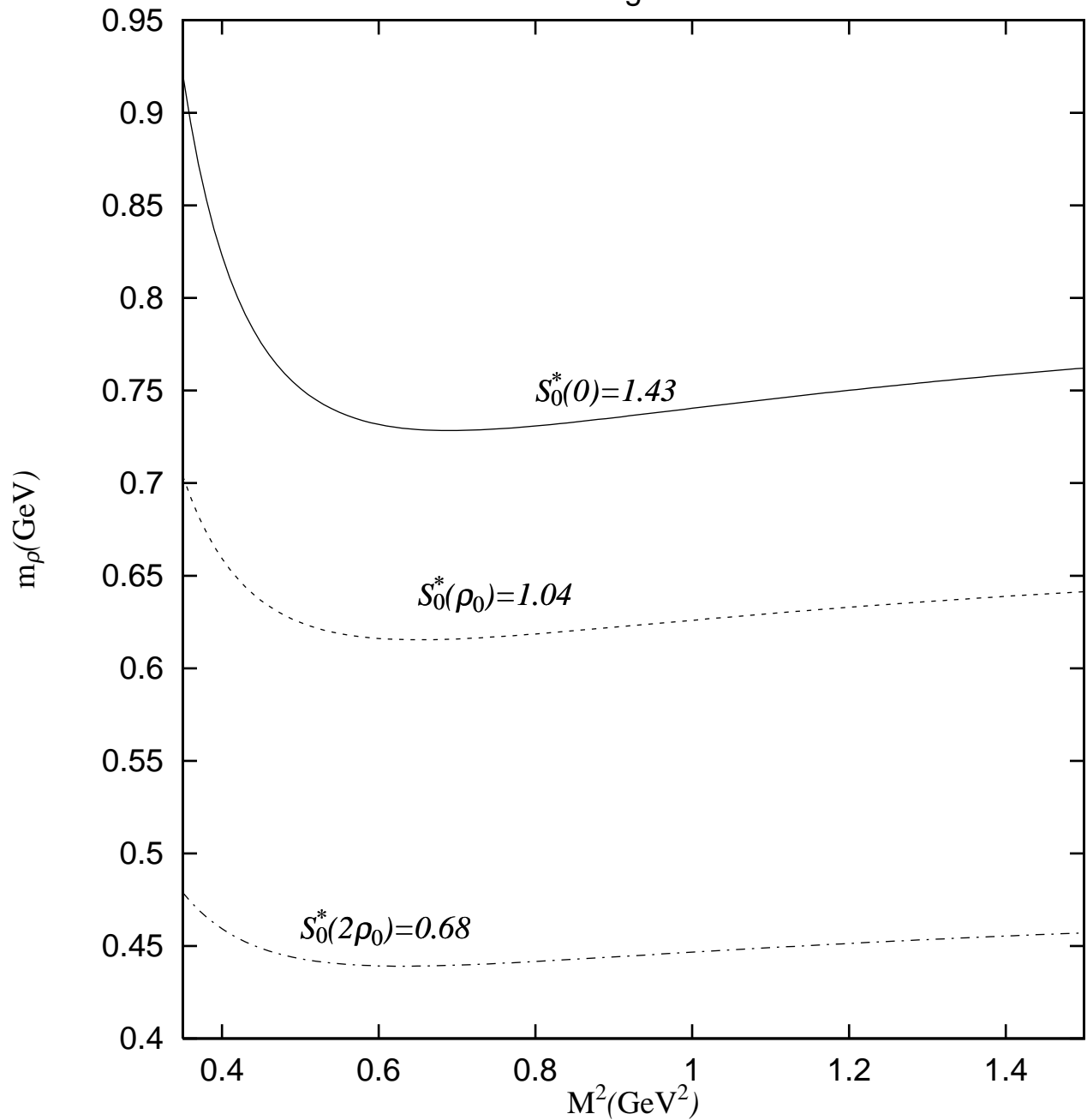
\item [Fig.2]
 Borel curve for the $\rho(\omega)$ meson mass.
 Solid, dashed and dash-dotted lines
 correspond to $\rho/\rho_0$ = 0, 1.0 and 2.0 respectively. $S_0^*(\rho)$ 
 determined by the
 Borel stability method at each density is also shown in GeV$^2$ unit.
The Borel window is chosen to be $0.41 {\rm GeV}^2 < M^2< 1.30 {\rm GeV}^2$. 

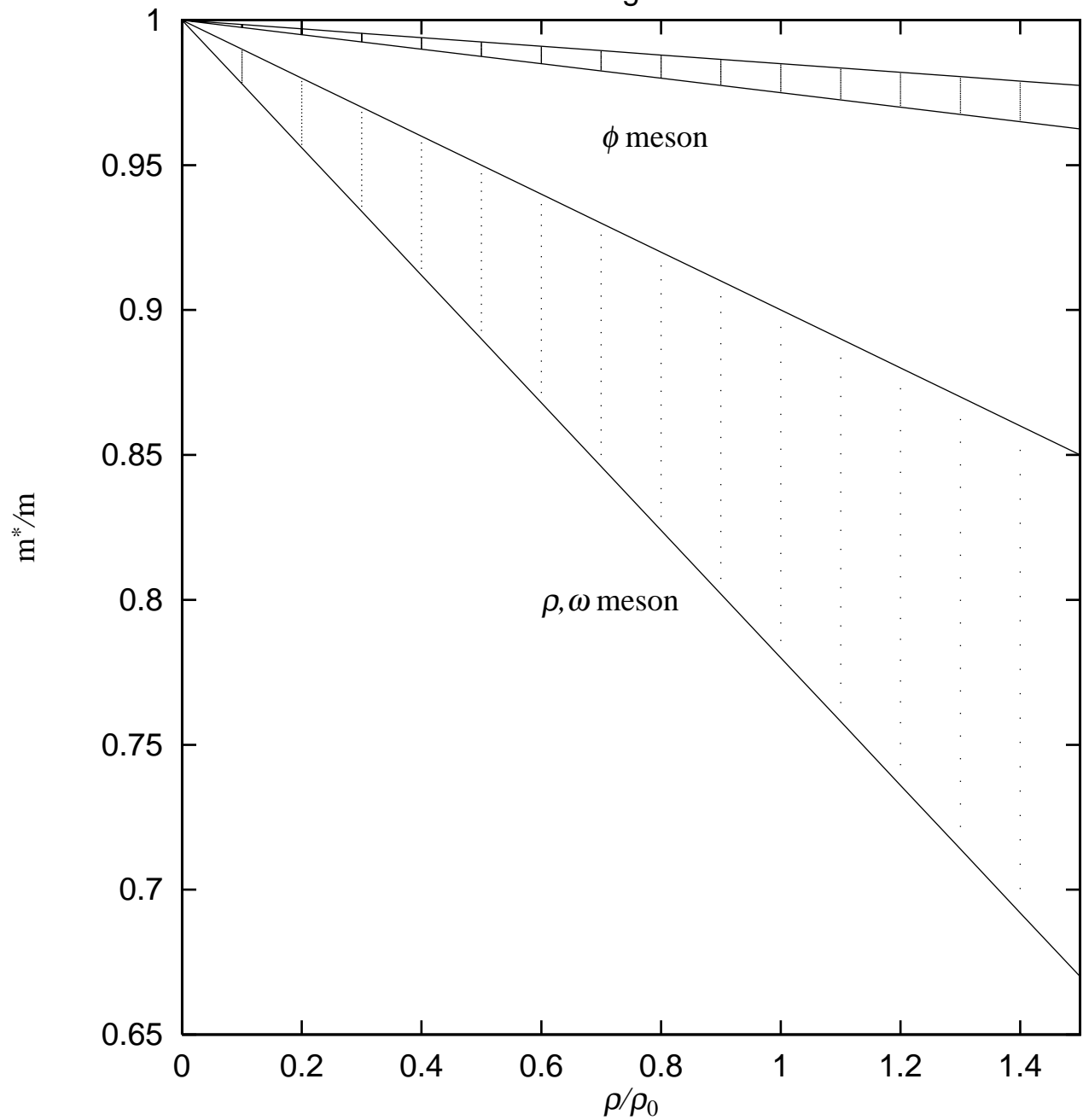
\item [Fig.3]
 Masses of $\rho, \omega$ and $\phi$ mesons 
 in nuclear matter predicted in the QCD sum rules.
  The hatched region shows  theoretical uncertainty.

\end{description}

\end{document}